\documentclass[sigconf,nonacm]{acmart}

\renewcommand\footnotetextcopyrightpermission[1]{}
\usepackage{booktabs}
\usepackage{graphicx}
\usepackage{subcaption}
\usepackage{amsmath}
\usepackage{hyperref}
\usepackage{float}
\usepackage[skip=10pt]{caption}

\AtBeginDocument{%
}

\title{Context-Aware Prediction of Student Quiz Performance with Multimodal Textbook Features}

\author{Samin Khan}
\affiliation{%
  \institution{Stanford University}
  \country{}
}
\email{samink@stanford.edu}

\begin{document}

\begin{abstract}
Educational platforms often predict student performance from prior interactions, but the assessment content itself also varies in linguistic and visual complexity. This paper studies whether lightweight content features extracted from CourseKata chapter-review questions improve prediction of end-of-chapter quiz scores beyond a student's average prior exercise performance. The study combines 2023 CourseKata student response data with chapter-level text features from review-question wording and image features from textbook visuals. Across 4,742 student-chapter observations from 562 class-student IDs, adding content features improves student-grouped five-fold quiz prediction performance by 9.1\% relative to a prior-performance baseline. In leave-chapter-out validation, text features reduce prediction error relative to the baseline, while image-containing models have higher error. This paper suggests that a context-aware model adds useful signal about the text and visual features of questions to better predict student quiz performance compared with using past student performance alone.
\end{abstract}

\keywords{
Learning Analytics,
Educational Data Mining,
Machine Learning,
Student Performance Prediction,
Feature Engineering
}

\maketitle

\section{Introduction}

Predicting student performance is a central task in educational data mining because accurate forecasts can support adaptive practice, formative feedback, and earlier identification of students who may need help. Classical knowledge tracing models estimate evolving student mastery from interaction histories \citep{corbett1995knowledge}, and modern neural variants have shown that flexible sequence models can capture richer patterns in student records \citep{piech2015deep}. However, a student's future score is not only a function of prior performance. The content of the target assessment can also vary in vocabulary, length, statistical topic, representational form, and visual complexity.

This distinction matters for online textbooks such as CourseKata, where students work through interactive statistics and data science materials before taking end-of-chapter quizzes. CourseKata provides large-scale anonymized data from college courses using the ABC and ABCD versions of \emph{Statistics and Data Science: A Modeling Approach} \citep{son2017coursekata}, whose pedagogical design emphasizes statistical modeling and connected conceptual understanding \citep{son2021modeling}. If assessment content varies in difficulty-relevant ways, then models that use only prior exercise scores may leave predictable variance unexplained.

The broad research question is therefore: \emph{Do text and image features from chapter-review content improve prediction of student quiz performance beyond prior exercise performance?} This question has two practically distinct forms. First, for student modeling in an existing course, the relevant question is whether content features improve prediction for held-out students when the same chapters are represented in the training data. In this setting, content features supplement prior student performance with information about how other students performed on similar chapter content. Second, for new-curriculum use cases, the relevant question is whether relationships learned from existing chapters transfer to a chapter withheld during training. This setting is closer to predicting quiz difficulty or expected performance for newly designed instructional content before large-scale student data are available.

Accordingly, this paper addresses two research questions:
\begin{enumerate}
  \item \textbf{RQ1: Held-out student prediction.} Do text and image features improve prediction of quiz scores for students not seen during training, when the target chapters are represented by other students in the training data?
  \item \textbf{RQ2: Held-out chapter prediction.} Do text and image features improve prediction of quiz scores for a chapter not seen during training?
\end{enumerate}

This paper answers these questions with an interpretable pipeline: aggregate prior exercise scores at the student-chapter level, extract lightweight text and image features from review questions, compare linear and regularized models, and evaluate both held-out-student and held-out-chapter generalization.

\section{Related Work}

Student performance prediction has increasingly moved from simple historical summaries toward models that incorporate richer exercise and item information. Exercise-aware knowledge tracing explicitly studies how exercise content can improve prediction and interpretability beyond interaction records alone \citep{liu2021ekt}. Recent multimodal work similarly argues that question semantics and student-question structure should be modeled together, including approaches that combine graph neural networks with language-model-derived question representations \citep{wang2024multimodal}.

This paper is also related to automatic item difficulty prediction. A systematic review of text-based question difficulty prediction found that linguistic features are frequently important and that automatic approaches can address the cost and scalability limits of expert calibration and pretesting \citep{alkhuzaey2023text}. Recent empirical work comparing text-feature models for item difficulty likewise shows that lexical and wording features can be useful predictors in assessment settings \citep{stepanek2023item}. The present study differs by predicting individual student quiz scores, not only item difficulty, while asking whether assessment content adds signal after controlling for each student's prior chapter performance.

\section{Data}

All data presented here were collected and shared with the author by CourseKata (\url{https://coursekata.org}). The analysis uses a processed split derived from the 2023 CourseKata college dataset. The end-of-chapter file contains student quiz scores for chapters 1--9, and the response file contains exercise-level attempts, points earned, and points possible. A student is represented as a unique class-student pair, consistent with CourseKata's public data documentation. After merging student exercise aggregates, quiz scores, and content features, the analytic dataset contains 4,742 student-chapter observations from 562 class-student IDs across five classes and two textbooks: ABC and ABCD.

The outcome is the end-of-chapter quiz score, scaled from 0 to 1. The baseline predictor is \texttt{avg\_exercise\_scores}, the mean proportion correct across prior exercises attempted by the same student in the same chapter.

\section{Methods}

\subsection{Content Feature Extraction}

Chapter-review pages were scraped from CourseKata preview pages using Selenium because the review items are dynamically rendered. The cleaned public code replaces the original notebook's unsafe \texttt{eval} parsing with \texttt{ast.literal\_eval} and separates scraping, feature construction, and modeling into reusable scripts.

Text features were computed with \texttt{wordfreq} \citep{wordfreq}: average word length, mean word rarity defined as $7 - \mathrm{ZipfFrequency}(w)$, number of complex words with Zipf frequency below 5, and total word count. These features approximate surface-level linguistic complexity, verbosity, and vocabulary rarity.

\begin{figure}[H]
  \centering
  \includegraphics[width=.85\linewidth]{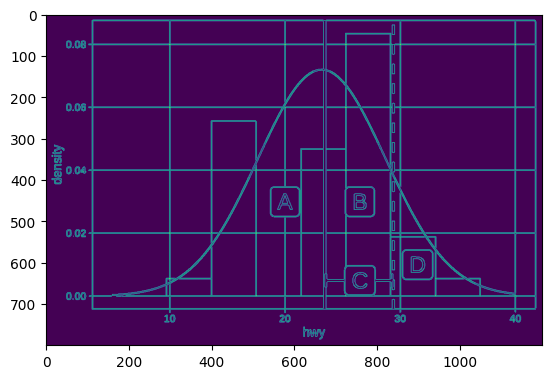}
  \caption{Example image associated with a CourseKata quiz question processed using Canny edge detection. 532 line segments and 58 visual regions were detected.}
  \Description{A processed textbook image showing edge-detected line and contour structure used to derive image-complexity features.}
  \label{fig:edge}
\end{figure}

Image features were computed with OpenCV \citep{opencv_library}. For each review-page image, Canny edges were extracted, Hough line segments were counted, and external contours were counted as a rough proxy for visual regions. Chapter-level image features are the average number of lines per image, average number of regions per image, and total image count. Figure~\ref{fig:edge} shows an example of this processing pipeline.

\subsection{Models and Evaluation}

The analysis compares four feature sets:

\begin{enumerate}
  \item \textbf{Base}: prior exercise performance only.
  \item \textbf{Base + Image}: base feature plus image features.
  \item \textbf{Base + Text}: base feature plus text features.
  \item \textbf{Base + Image + Text}: all features.
\end{enumerate}

For comparability with the original notebook, ordinary least squares models are first fit in sample. The primary evaluation for RQ1 is five-fold grouped cross-validation where all chapters from the same class-student ID are kept in the same fold. In each split, the model is trained on four folds of students and tested on the held-out fold of students. The chapters are represented in both train and test folds, but no class-student ID appears in both. This evaluates generalization to new students in known course content.

The evaluation for RQ2 is leave-chapter-out validation. In each split, the model is trained on all observations from eight chapters and tested on all observations from the remaining chapter. This evaluates whether content-feature relationships learned from existing chapters transfer to an unseen chapter. Ridge regression was used because the content features are correlated and regularization helps reduce overfitting; the regularization strength was selected by cross-validation within each training fold.

\section{Results}

\subsection{RQ1: Held-Out Student Prediction}

Table~\ref{tab:main_results} reports the student-grouped cross-validation results. The baseline model reaches $R^2=0.3775$ and MSE $=0.02496$. Adding image features yields MSE $=0.02425$. Adding text features yields MSE $=0.02275$. The full multimodal model yields $R^2=0.4338$ and MSE $=0.02270$, a 9.1\% relative MSE reduction compared with the baseline. Statistical significance was assessed with a paired bootstrap over class-student IDs: for each augmented model, each class-student ID's mean squared-error difference versus the baseline was computed, these grouped differences were resampled 5,000 times with replacement, and a two-sided bootstrap p-value with a 95\% percentile confidence interval was used. Asterisks in Table~\ref{tab:main_results} mark augmented models whose MSE improvement over the baseline was significant at $p<.05$ and whose 95\% confidence interval excluded zero.

\begin{table}[htbp]
\centering
\begin{tabular}{lccc}
\toprule
Model & $R^2$ & MSE & MSE reduction \\
\midrule
Base & 0.3775 & 0.02496 & -- \\
Base + Image* & 0.3953 & 0.02425 & 2.9\% \\
Base + Text* & 0.4326 & 0.02275 & 8.9\% \\
Base + Image + Text* & \textbf{0.4338} & \textbf{0.02270} & \textbf{9.1\%} \\
\bottomrule
\end{tabular}
\caption{Student-grouped five-fold cross-validation results. Quiz performance prediction error decreases when text and/or visual features are incorporated into the model ($p<.05$).}
\label{tab:main_results}
\end{table}
\subsection{RQ2: Held-Out Chapter Prediction}
Table~\ref{tab:chapter_results} reports the leave-chapter-out validation results. The baseline model reaches $R^2=0.3562$ and MSE $=0.02582$. The base + text model yields $R^2=0.4119$ and MSE $=0.02358$, an 8.7\% relative MSE reduction compared with the baseline. The base + image model yields $R^2=0.2564$ and MSE $=0.02982$. The full multimodal model yields $R^2=-0.1776$ and MSE $=0.04722$.

\begin{table}[H]
\centering
\begin{tabular}{lccc}
\toprule
Model & $R^2$ & MSE & MSE reduction \\
\midrule
Base & 0.3562 & 0.02582 & -- \\
Base + Image & 0.2564 & 0.02982 & -15.5\% \\
Base + Text & \textbf{0.4119} & \textbf{0.02358} & \textbf{8.7\%} \\
Base + Image + Text & -0.1776 & 0.04722 & -82.9\% \\
\bottomrule
\end{tabular}
\caption{Leave-chapter-out validation results. Quiz performance prediction error decreases when text features are incorporated into the model, whereas incorporating image features increases error.}
\label{tab:chapter_results}
\end{table}

\section{Discussion}

The results support two main findings. First, chapter-review content contains predictive information beyond prior exercise performance for held-out students in known curriculum. This is the setting most closely aligned with operational student-performance prediction: the model has access to how other students performed on the same chapters, and content features add information beyond each student's chapter-specific exercise history.

Second, the held-out-chapter results distinguish text features from the current image features. Text features reduce MSE relative to the baseline when the target chapter is excluded from training, which suggests that chapter-level linguistic features capture some information that transfers across curriculum units. This aligns with prior item-difficulty research showing that linguistic features often help explain assessment difficulty \citep{alkhuzaey2023text,stepanek2023item}. In contrast, models that include the current image features have higher leave-chapter-out error than the baseline. This leaves room for feature-engineering work on models intended to predict performance for newly designed curriculum or assessment content.

\begin{figure}[H]
  \centering
  \includegraphics[width=.78\linewidth]{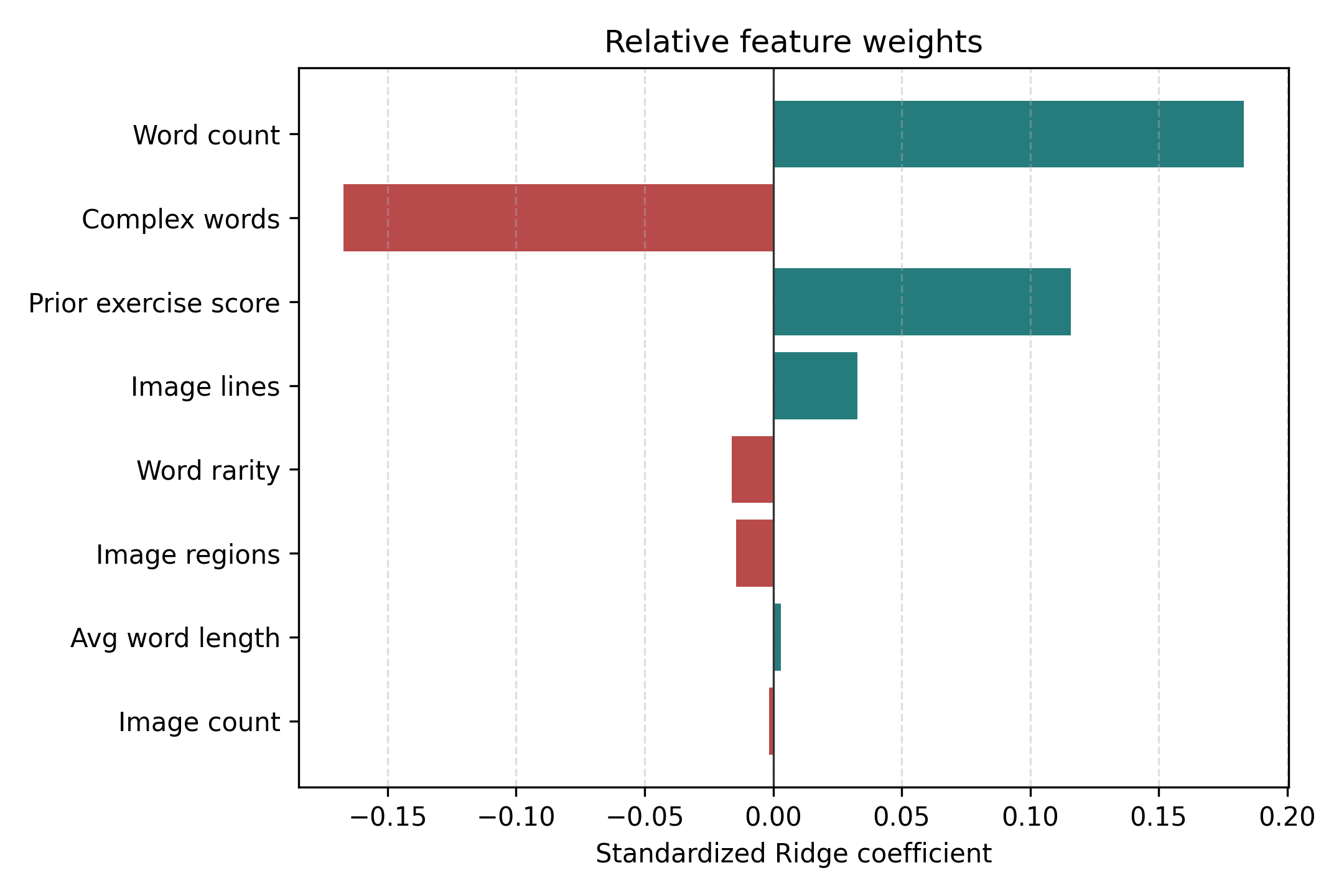}
  \caption{Standardized Ridge coefficients for the full model. Text-count features carry the largest content-based weights, while most image features are smaller.}
  \Description{Horizontal bar chart of standardized Ridge coefficients. Word count, complex words, and prior exercise score have the largest magnitudes.}
  \label{fig:weights}
\end{figure}

Standardized coefficient inspection of a Ridge model fit with all features provides descriptive context for the source of the prediction signal. Figure~\ref{fig:weights} shows that word count and complex-word count have the largest content-feature weights, followed by the prior exercise score and smaller image and lexical-rarity terms. Because the features are correlated and the coefficients are not causal effects, the analysis should be interpreted as evidence about the relative predictive role of the engineered features rather than as an explanation of question difficulty.

The image-feature results should be interpreted in light of the simplicity of the visual representation. Edge counts, contour counts, and image counts are coarse proxies for visual demand. A chapter with many lines may contain a complex statistical graphic, a clean table, or a screenshot-like artifact; the feature does not encode which visual relations the student must interpret. Future work should replace or supplement these features with semantically meaningful representations such as chart type, axis clarity, number of encoded variables, or vision-language embeddings.

\section{Limitations}

First, the content features are chapter-level aggregates, not item-level features linked to each quiz question a student answered. This limits precision and may blur differences among questions within a chapter. Second, the models are intentionally simple and interpretable; they do not capture nonlinear interactions between student history and content complexity. Third, the analysis uses a subset of CourseKata classes from one assignment split rather than the entire public CourseKata release. Fourth, although the content features improve prediction, they should not be interpreted as causal estimates of what makes a question difficult.

\section{Conclusion}

This study shows that lightweight context-aware features can improve prediction of CourseKata end-of-chapter quiz scores beyond prior exercise performance. In held-out-student prediction, text and image features reduce prediction error relative to a prior-performance baseline. In held-out-chapter prediction, text features reduce prediction error, while the current image features do not. The broader implication is that student modeling should treat assessment content as part of the prediction problem, while using evaluation designs that distinguish known-content personalization from new-content generalization.

\section*{Code Availability}

The accompanying repository includes cleaned scripts for feature construction and model evaluation, a feature-only processed table, and reproducible output generation.
\url{https://github.com/samin-khan/predicting-student-quiz-performance-from-textbook-features}

\bibliographystyle{ACM-Reference-Format}
\bibliography{references}

\end{document}